\documentclass[]{spie}  

\usepackage{amsmath,amsfonts,amssymb}
\usepackage{graphicx}
\usepackage[colorlinks=true, allcolors=blue]{hyperref}
\usepackage{tikz}
\usetikzlibrary{arrows.meta,positioning,calc,fit,backgrounds}
\usepackage{adjustbox}
\usepackage{xcolor}
\usepackage{subcaption}
\usepackage{url}

\usepackage{siunitx}

\usepackage{makecell}

\usepackage{array} 
\usepackage{booktabs} 
\usepackage{arydshln} 

\usetikzlibrary{math}

\DeclareSIUnit{\sqrthz}{\ensuremath{\sqrt{\text{\hertz}}}}

\title{picoMUX: microcontroller based time-domain multiplexing readout for kilopixel TES arrays}

\author[a]{Simon~Tartakovsky}
\author[b]{Alexandre~E.~Adler}
\author[c]{Jason~E.~Austermann}
\author[a]{Steven~J.~Benton}
\author[c]{Shannon~M.~Duff}
\author[g,c]{Malcolm~Durkin}
\author[e]{Jeffrey~P.~Filippini}
\author[a]{Aurelien~A.~Fraisse}
\author[f]{Thomas~J.L.J.~Gascard}
\author[e]{Sho~M.~Gibbs}
\author[a]{Suren~Gourapura}
\author[f,~b]{Jon~E.~Gudmundsson}
\author[c]{Johannes~Hubmayr}
\author[a]{William~C.~Jones}
\author[f]{Ashesh~Khatua}
\author[d]{Jared~L.~May}
\author[e]{Darby~McCauley}
\author[d]{Johanna~M.~Nagy}
\author[d]{Ivan~L.~Padilla}
\author[d]{Ricardo~R.~Rodriguez}
\author[d]{John~E.~Ruhl}
\author[d]{M.~Shaaf~Sarwar}
\author[c]{Christopher~Rooney}
\author[d]{Joseph~van~der~List}
\author[c]{Michael~R.~Vissers}
\author[a]{Philippe~Voyer}

\affil[a]{Department of Physics, Princeton University, Jadwin Hall, Princeton, NJ 08544, USA}
\affil[b]{The Oskar Klein Centre, Department of Physics, Stockholm University, AlbaNova, SE-10691 Stockholm, Sweden}
\affil[c]{National Institute of Standards and Technology, 325 Broadway Mailcode 817.03, Boulder, CO 80305, USA}
\affil[d]{Department of Physics, Case Western Reserve University, 10900 Euclid Ave, Cleveland, OH 44106, USA}
\affil[e]{Department of Physics, University of Illinois Urbana-Champaign, 1110 W Green St, Urbana, IL 61801, USA}
\affil[f]{Science Institute, University of Iceland, 107 Reykjavik, Iceland}
\affil[g]{Department of Physics, University of Colorado Boulder, Boulder, Colorado, USA}

\authorinfo{Further author information: (Send correspondence to S.T.)\\S.T.: E-mail: simont@princeton.edu}

\begin{document} 
\maketitle

\renewcommand{\arraystretch}{1.15}

\newcommand{\st}[1]{\textcolor{red}{#1}}

\begin{abstract}
Advances in cryogenic SQUID-based time-domain multiplexing (TDM) have outpaced their warm readout electronics. 
Next generation CMB telescopes are baselining kilopixel TES arrays with few viable electronic options. 
We present picoMUX, a new TDM readout electronics architecture that replaces the FPGA typically used in such systems with a complement of modern microcontrollers.
Designed around the current generation NIST multiplexer, it fully exploits its differential nature and two-level switching.
The system achieves the desired timing metrics (sub \unit{\micro\second} row dwell time) while reducing cost, complexity, and power consumption over a comparable FPGA solution. 
Preliminary noise measurements are consistent with expectations and show no evidence that the novel architecture introduces excess noise. 
picoMUX demonstrates how recent advances in microcontrollers enable simpler, low-power, low-cost TDM readouts for kilopixel TES arrays.
\end{abstract}


\section{Introduction}
\label{sec:intro}

Next generation cosmic microwave background (CMB) experiments require kilopixel sized transition-edge sensor (TES) arrays to achieve their sensitivity goals\cite{taurus_overview,s4_techbook}.
The mature SQUID based time-domain multiplexing (TDM) scheme remains an attractive readout approach thanks to continued development\cite{nist_squids,s4_readout}.
TES multiplexing reduces the amount of cryogenic wiring at the cost of readout complexity. 
For TDM this translates to difficult timing requirements for the warm readout.

Existing TDM warm readout systems tackle the timing complexity by using field-programmable gate arrays (FPGAs)\cite{Battistelli:2008skh}. 
FPGAs are well suited for this application and are able to fulfill the requirements by parallelizing tasks such as analog-to-digital sampling, digital filtering, and feedback computation.
The flexibility of FPGAs comes at the cost of increased power consumption and cost compared to purpose built processors.
It also increases the complexity of programming and debugging, frequently to the point that the scientists working with them day-to-day are no longer able to make meaningful improvements.

This work presents the development of picoMUX, a microcontroller-based TDM readout system designed to work with kilopixel sized TES arrays.
picoMUX leverages the fact that the central difficulty of TDM is not the complexity of the feedback calculation but rather the small number of input/output operations that must happen at deterministic times within the sub-microsecond readout window.
These strict timing windows are satisfied by making full use of the rich peripheral set of the RP2350 microcontroller to achieve FPGA-like performance without the associated complexities.

To motivate the architecture of picoMUX, Section~\ref{sec:tdm} reviews the operational principal of TDM with a focus on timing.
Section~\ref{sec:architecture} presents the overall readout architecture and its mapping onto the RP2350 microcontroller. 
Finally, Section~\ref{sec:perf} discusses the performance metrics of picoMUX and presents preliminary measurements in a 4\,K development cryostat.

\section{Timing requirements of TES Time-domain multiplexing}
\label{sec:tdm}

In time domain multiplexing, a set of $N_{\rm row}$ TES detectors is read out sequentially within a frame. 
The frame rate, or revisit rate, $f_r$ sets the Nyquist bandwidth of the detectors and must be high enough to capture the full response of the TESs without unacceptable aliasing of signal and noise.
Since all rows must be read out within one frame, the switching frequency is $f_s= N_{\rm row} f_r$ and the corresponding row dwell time is $t_{row} = 1/f_s$.
The row dwell time is the natural quantity used for timing requirements as all row-level operations must occur within this interval.

\begin{figure}
    \centering
    \includegraphics[width=0.7\textwidth]{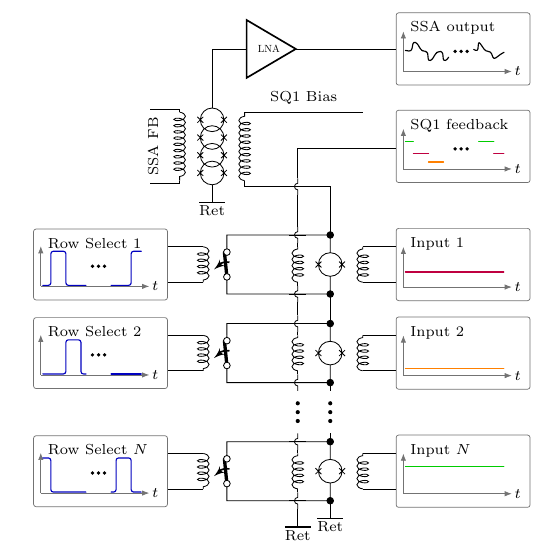}
    \caption{Schematic diagram of one column of a time domain multiplexing (TDM) system. The series SQUID array (SSA), depicted by a stack of 4 SQUIDs, is connected to a set of $N_{\rm row}$ SQUIDs (SQ1s), each shunted by a normally closed flux activated switch (FAS). A sequence of row-select pulses open a single FAS at a time to bias a specific SQ1 associated with a color coded input. Current generation TDM chips use two-level switching and include chip select FASs that bypasses an entire bank of SQ1s to further reduce wiring. The SQ1 feedback signal computed on the previous row visit nulls the input signal, keeping the SSA output near its setpoint. The SSA output shows typical switching transients followed by a settling, set by the bandwidth of the readout chain. Biasing resistors, fully differential return lines, and the SSA bias line, have been omitted from the diagram for clarity.}
    \label{fig:tdm_diagram}
\end{figure}

The typical layout of a TDM column is shown in Fig.~\ref{fig:tdm_diagram}.
Each TES is coupled to a first-stage SQUID, denoted SQ1, through a dedicated input coil.
The SQ1s are connected in series and all share a single bias line that passes through the input coils of a SQUID series array (SSA) for additional amplification before being digitized by the warm electronics.
To select a particular TES, each SQ1 is shunted by a flux activated row-select switch (FAS) that is normally closed.
These row-selects can be asserted sequentially by a control current, allowing the readout system to visit each TES in the column row by row\cite{nist_squids}.

The SQUIDs that make up the cold readout exhibit excellent noise performance, but due to their non-linear response, must be operated in a flux locked loop (FLL).
In FLL operation, the readout system drives the feedback coils to keep all the SQUIDs in the readout chain near a fixed operating point.
The detector signal is then inferred from the feedback value required to null the SQUID response, rather than from the open-loop SQUID output.
In a TDM column, the controller must maintain an independent feedback state for each TES, even though only one row is actively read out at a time. 
These feedback states are updated in a time-interleaved manner as the row-select advances through the frame.

The sequence of operations during one row dwell is highly structured.
At the start of the row visit, the controller outputs the stored feedback value and biasing parameters\footnote{Examples include per-SQ1 bias and SSA feedback. Current generation NIST TDM devices have improved uniformity and reduced the need for unique biasing parameters for each SQ1. Nevertheless, picoMUX maintains the ability to output per row SSA feedback but is limited to a single SQ1 bias per column.} associated with the current row simultaneously with the application of the appropriate row-select current.
After a short settling period, the SSA error signal is sampled by the analog-to-digital converter (ADC).
The controller computes an updated feedback value for that row, stores it for use in the next frame, and advances to the next row.
The resulting feedback values track the input signals and are output from the readout system after being low pass filtered and decimated.

Typical bolometric applications require the revisit rate to be $\approx$10\,\unit{\kilo\hertz} to avoid aliasing detector noise.
Combined with the desired multiplexing factors of $\approx$80, the required row visit time is around 1\,\unit{\micro\second}\cite{ade2014bicep2}.
The challenge of TDM readout electronics is to run all these concurrent FLLs in a power efficient manner while only using less than a microsecond of compute time per iteration. 

In terms of raw computation, a TDM system needs only to compute a single feedback value and run a decimating digital filter at each row visit.
Even at switching frequencies above 1\,MHz, the corresponding arithmetic throughput is well within the capability of modern embedded processors. 
The stringent requirement is rather the deterministic scheduling of the digital-to-analog converter (DAC) updates, row-select transitions, and ADC sampling that must occur at precise offsets within each row dwell window.

The distinction between computational load and timing-sensitive input and output is what makes picoMUX possible. 
Rather than using an FPGA as a general purpose controller that can accomplish both these tasks, picoMUX maps the deterministic parts of TDM onto microcontroller peripherals, while reserving the processor cores for the arithmetic load that they were optimized for.
picoMUX does not relax the timing requirements of TDM, instead it splits the difficulty into two parts, a real time interface task and a computational problem, and assigns them to the appropriate resource available in a modern microcontroller.

\section{System Architecture}
\label{sec:architecture}

picoMUX uses a distributed architecture composed of readout cards and address cards interfaced with the user over Ethernet.
The readout cards contain the hardware required to manage up to eight TDM columns, while the address card manages synchronization and generates row-select waveforms for up to 16+16 row and chip selects.
These cards connect directly to the cryostat flanges using pinouts compatible with the standard established by the Multi-Channel Electronics (MCEs) developed by The University of British Columbia (UBC).

The basic building block of picoMUX is the column module.
On the readout card, each column module manages a single TDM readout column: it runs the interleaved FLLs for each row by sampling the SSA and outputting to the SQ1 feedback line.
On the address card, this module, with additional external components, is reused to generate row-select waveforms.

\subsection{Column Module}
\label{sec:column_module}

\begin{figure}
    \centering
    \includegraphics{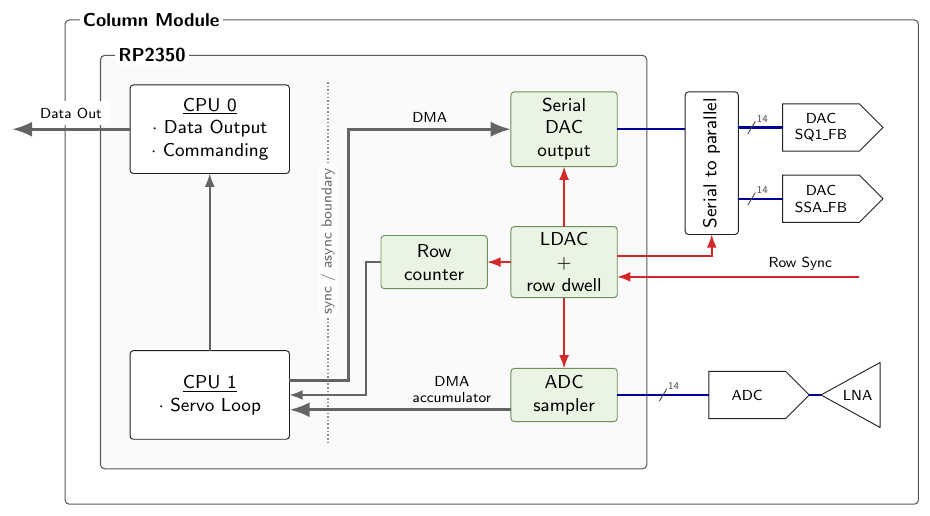}
    \caption{Simplified block diagram of a column module showing hardware and software architecture. Red arrows indicate hardware triggers. Green boxes represent PIO blocks that are used to decouple critical timing from the CPU. The system uses two fast DACs for SQ1 and SSA feedback so the FLL can be closed either on the SQ1 or SSA for tuning, as shown in the left panel of Fig.~\ref{fig:noise_plots}. Bias DACs, clocking, power supplies, and address card interconnects are not shown. Each readout card has a low power processor that aggregates data from the 8 column modules and relays it to the user over Ethernet. }
    \label{fig:block_diagram}
\end{figure}

The column module, depicted in Fig.~\ref{fig:block_diagram}, pairs a single RP2350 microcontroller with the hardware required to drive a single TDM readout column.
The RP2350 is well suited for the task because, in addition to the dual ARM cores, it features programmable input/output (PIO) peripherals.
These PIOs are small state machines with deterministic timing and extremely reduced instruction sets designed for high speed digital interfaces.
picoMUX uses these peripherals to offload timing-critical TDM input and output tasks from the processor cores.

picoMUX is designed such that all deterministic row-level operations are performed by peripherals, while the processor cores only run the feedback and filtering operations.
This separation prevents timing jitter associated with instruction execution on complex processors from coupling directly into the strict timing deadlines of TDM.
The RP2350 is not used primarily as a conventional sequential microcontroller, but as a collection of deterministic peripherals coordinated by the processor cores.
This allows for the parallelization of TDM tasks, analogous to the way in which such a system would be implemented on FPGAs.

The module contains a low-noise analog front end, a 50\,MS/s analog to digital converter (ADC) used to sample the SSA, and two high-speed digital-to-analog converters (DACs)  used to drive the SQ1 and SSA feedback lines.
The analog design of picoMUX draws from the heritage of the MCEs, but is adapted for current generation TDM readout. It makes use of newer high performance components, optimized for power consumption, and integrated with the picoMUX digital scheme.
The selected ADC for this application outputs data over a 14 bit parallel interface.
This interface is directly compatible with the PIO blocks, which include a parallel load instruction capable of sampling digital busses.
By pairing the PIO with the direct memory access (DMA) engine, picoMUX samples the ADC and applies a boxcar filter without intervention from the main processor.

A second PIO/DMA in combination with a serial-to-parallel converter is used to output the SQ1 feedback and SSA feedback DAC codes.
After receiving a synchronization pulse from the address card, the PIO and DMA load the appropriate feedback values and update the DAC at deterministic offsets relative to the row timing.
In fact, all timing critical tasks are performed using PIO and DMA to guarantee that the strict timing requirements outlined in section~\ref{sec:tdm} are adhered to.

With all timing-critical operations handled by peripherals, the processor cores attend to the remaining tasks of feedback computation and data filtering.
One of the CPU cores waits on the PIOs, that, upon completing a row visit, indicate the row number that was visited along with the accumulated ADC signal, allowing the feedback computation to begin.
These computations must maintain the same throughput as the rest of the system, but since the resulting feedback is only required for the subsequent frame (tens of \unit{\micro\second}), some timing uncertainty can be accommodated.
The feedback calculation is simple and well suited for a CPU but requires the core to access many values from memory (gain, setpoint, etc...), adding substantial overhead.
Minimum row dwell is set by the amount of time the core requires to run this computation.
A complete timing diagram for picoMUX is shown in Fig.~\ref{fig:timing_diagram}.

The final challenge for picoMUX is to get the data stream out of the microcontroller and onto persistent storage. 
Nominally, picoMUX would filter and decimate the data to match the expected detector bandwidth, resulting in data rate of only a few hundred kB/s per column.
During system development and debugging, it is useful to access the full-rate diagnostic data that can exceed 10\,MB/s and overwhelm most common peripherals. 
To support this mode, picoMUX uses an additional PIO block to emulate a parallel digital camera style interface capable of sustaining the increased data rate. 

\subsection{Address Card}
\label{sec:addres_card}

The address card generates the row-select waveforms and provides the global timing reference for the picoMUX system.
Unlike the readout card that is simply a carrier for up to 8 column modules, the address card contains active circuitry that routes high-speed DAC output to the row-select lines.

To reduce complexity, the address card reuses a single column module.
The two high-speed DACs on the module are connected to two banks of analog multiplexers, allowing the module to drive up to 32 row-select outputs.
The same digital data bus used for data output on the readout card is repurposed as an address bus for selecting the active multiplexer output. 
In this configuration, the column module configures the multiplexer to route the appropriate signal to the desired row-select line.

picoMUX uses two banks of multiplexers so that two row-select outputs can be driven simultaneously.
These 1-16 multiplexers can be used to either drive a modern two level-switching readout with one bank acting as row-select and the other as chip-select, or they can be used independently to drive a 32 row single-level system.

Since typical picoMUX systems will have one address card and many readout cards, the address card generates all timing pulses used to synchronize the cards.
The beginning of a new frame is indicated by a \texttt{row0\_sync} pulse and the boundaries between row dwell intervals are marked by \texttt{row\_sync} pulses.
These synchronization signals, routed directly to all readout cards, are used by the PIO blocks to align DAC updates, ADC sampling, and feedback indexing to the active row.

\section{Performance}
\label{sec:perf}

The performance measurements of picoMUX are not intended to demonstrate a fundamentally new readout chain. 
The analog design is intentionally conservative and draws heavily from the heritage of the MCE system, reusing multiple key components. 
A row-select multiplexer, while not used by the MCEs, has been demonstrated in other TDM readout systems\cite{rs_multiplexer}.
The central performance question for picoMUX is therefore whether the new digital architecture can meet the timing requirements of TDM while reducing the power and complexity of the digital section.
This section focuses on the feedback-computation margin, and digital-section power consumption. 
Preliminary noise measurements are also presented to verify that the architectural changes in picoMUX do not compromise the expected readout noise performance.

\subsection{Digital Performance}
\label{sec:digital_perf}

FPGA-based TDM solutions can parallelize and pipeline digital tasks so that, in practice, the minimum row dwell time is dictated by analog bandwidth of the readout chain.
picoMUX on the other hand, relies on some serial processing and therefore there is a hard limit on the minimum row dwell time that can be achieved.

\begin{figure}
    \centering
        \includegraphics[width = 1.0\textwidth]{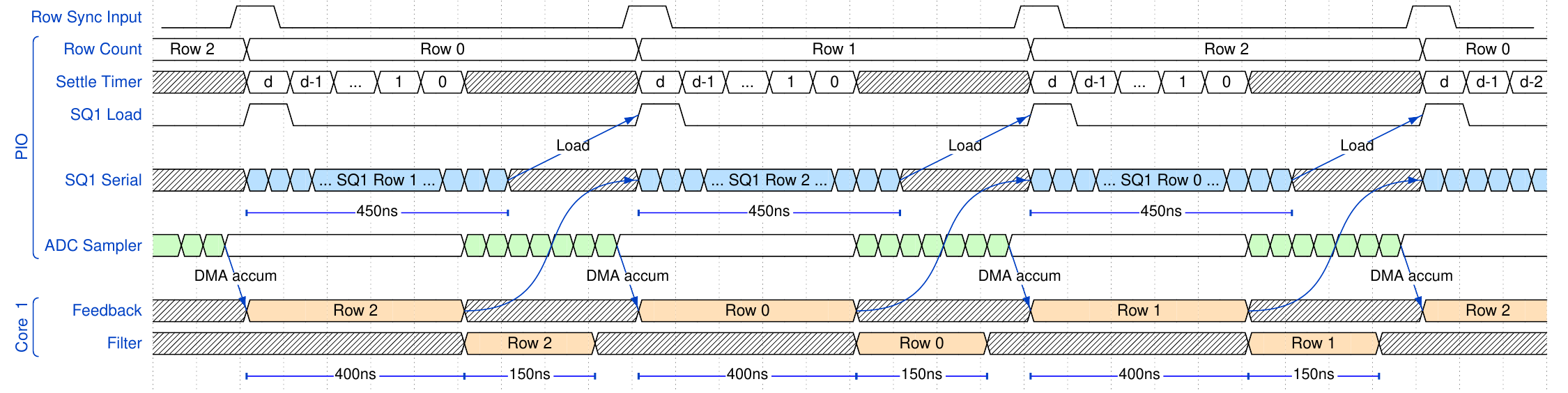}
    \caption{Timing diagram of a picoMUX system with 3 rows. A readout frame is initiated by an external row sync pulse. Synchronous operations, such as loading SQ1 codes and sampling SSA data are done with deterministic PIOs. Once the SSA data is collected and accumulated, the core computes the feedback value used for the subsequent row visit. Measured timings of serial tasks are presented to show the minimum row dwell time that can be achieved.}
    \label{fig:timing_diagram}
\end{figure}

The feedback computation run on the processor core is the limiting factor for row dwell, setting a bound of around 600\,ns.
If required, more performance can be achieved by micro-optimizing the code at the cost of simplicity and maintainability.
However, since the obtained switching time is well within the bounds of typical applications set by the analog bandwidth of the SQUID amplifier chain, such optimizations were not pursued.
A detailed breakdown of the timing is shown in Fig.~\ref{fig:timing_diagram}.

The main processor of the column module, the RP2350, consumes 50\,\unit{\milli\watt} to drive the FLLs for a single TDM column.
As a point of comparison, the digital section of the MCE system requires roughly $\approx$500\,\unit{\milli\watt} per column for the FPGA to achieve the same goals\cite{MCE_power}.
While keeping all analog components identical, switching from the traditional FPGA based digital architecture system to the architecture proposed by picoMUX can save full-scale deployments hundreds of watts.
Stratospheric balloon based experiments, such as Taurus, are extremely constrained on their power budgets and therefore would benefit greatly from such power savings\cite{taurus_overview}. 

The picoMUX system consumes roughly 1\,W per readout column, saving around a factor of 2 over the MCE, thanks to the negligible power consumption of the digital section and power conscious component selection in the analog chain.
A full scale deployment such as Taurus, composed of 192 columns, will consume under 200\,W as compared to the $\approx$500\,W required by the MCE to achieve the same channel count. 

\subsection{SQUID Tuning and Noise}
\label{sec:squid_perf}

\begin{figure}
\centering
\begin{subfigure}{.5\textwidth}
  \centering
  \includegraphics[width=.9\linewidth]{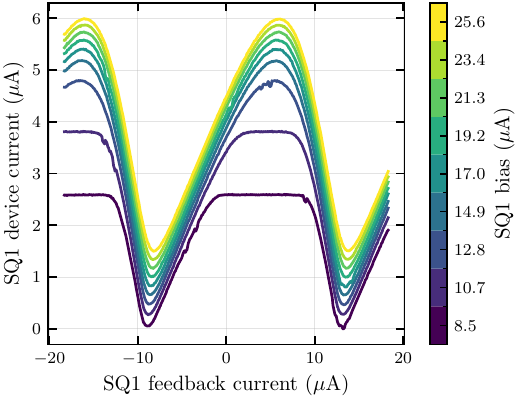}
\end{subfigure}%
\begin{subfigure}{.5\textwidth}
  \centering
  \includegraphics[width=.9\linewidth]{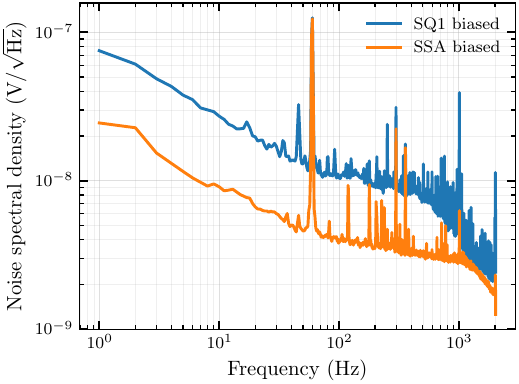}
\end{subfigure}
\vspace{0.1cm}
\caption{\emph{Left}: SQ1 tuning plot for a single SQ1 showing the response, with arbitrary zero, to a feedback ramp at different bias levels. The FLL is closed on the SSA during this procedure. The tuning procedure is optimized for throughput and therefore narrowband RF interference present in the lab environment is not averaged away and show up as small disturbances in the curves. \emph{Right}: Output referred noise at the SSA showing open loop noise of picoMUX multiplexing a single row. The measured white noise level for the SSA and SQ1 is within expectations for a system running at 4\,K. See Section~\ref{sec:squid_perf} for further discusion.}
\label{fig:noise_plots}
\end{figure}

To begin multiplexing an array of TESs, the operating points of the SQUIDs in the TDM chain must be determined.
The procedure for choosing the SQUID biases, feedback setpoints, and FLL gains is referred to as tuning and has been extensively discussed in the literature \cite{henderson2016readout, switch_tune}.
A set of automated scripts has been developed for picoMUX to tune a full array in under a minute.
The final step in the tuning process, indicating that the full chain is working as intended, is a SQ1 sweep, presented in the left panel of Fig.~\ref{fig:noise_plots}.
In this stage of tuning, the FLL is closed on the SSA feedback coil while the SQ1 feedback is ramped through a flux quantum to see the full response of the SQUIDs.
For time efficiency, all SQ1s are tuned simultaneously by multiplexing over all rows with similar parameters to those used when reading out the full array.
The presented SQ1 tuning plot therefore not only demonstrates that picoMUX can tune a TDM array of SQUIDs, but also the ability to multiplex and run the required concurrent FLLs.

picoMUX has been developed in a retrofitted cryostat that was never designed for low-noise device characterization.
This cryostat only reaches a base temperature 4\,K and does not have proper RF mitigation, such as dedicated shielding or grounding.
As a result, the system is particularly sensitive to interference, and due to the increased temperature, the SQ1 noise floor is not directly comparable to the design specification.
In addition, the elevated base temperature means that no TESs can be tested. 
Instead there are 100\,$\Omega$ resistors placed across the input coils of the SQ1s.
The right panel of Fig.~\ref{fig:noise_plots} presents the open loop SSA output referred noise for a biased SSA and SQ1 when multiplexing a single row.

The baseline SSA power spectrum shows performance roughly in line with the expectation for this device.
The prominent peaks in both traces are 60\,Hz mains pickup, harmonics, and other sources of RFI present in the lab.
The $1/f$ knee at around 10\,Hz is in line with the expectations for the amplifiers used in the frontend.
Biasing the SQ1 increases the noise floor, particularly at low frequencies, as it greatly increases the sensitivity of the system to RF interference.
The SQ1 timestream was pre-processed by removing transient glitches prior to estimating the power spectrum.
Since the SQ1s are designed to operate at temperatures under 500\,\unit{\milli\kelvin}, the resulting noise spectrum is not directly comparable to the design specification.
However, similar tests at elevated temperature performed at NIST show that the measured performance is not inconsistent with expectations. 
The clean SSA-biased power spectrum suggests that the additional RF sensitivity observed when the SQ1 is biased is specific to the test cryostat and lab environment rather than intrinsic to the picoMUX system. 

Typical TES bolometers optimized for CMB will exhibit a noise equivalent current of $\approx$100\,\unit{\pico\ampere/\sqrthz}.
Combining with the typical SQ1 gain of $\approx$4\,\unit{\micro\ampere/\micro\ampere} and SSA transimpedance to refer to SSA output results in an expected noise density of $\approx$50\,\unit{\nano\volt/\sqrthz}.
Even in this 4\,K development system, the noise measurements show that the $\sqrt{N_{\rm row}}$ noise penalty would still allow for multiplexing tens of rows before the aliased SQUID noise exceeded the intrinsic TES noise.
Notably, since multiplexing increases the noise floor by aliasing, the $1/f$ character of the noise will remain the same while the white noise level increases, resulting in an apparent reduction of the $1/f$ knee.

One method for mitigating $1/f$ noise in TDM readouts that involves modulating the input signal by alternating lock points was proposed by K. Sakai \cite{1/f_mitigation}.
picoMUX's architecture is uniquely well suited to attempt new feedback schemes, as the servo loop is written in pure C and can be recompiled and flashed in a few seconds.
This simplicity makes complex debugging and introspection of the system accessible, as all internal state variables can be outputted with a fast re-compilation of the code.

\section{Conclusion}

We present the architecture and preliminary performance for the picoMUX TDM readout system.
picoMUX demonstrates that advances in modern microcontrollers can be used to tackle problems that have traditionally been only accessible by FPGAs.
For TDM, switching to an FPGA-free architecture results in cost and power savings, as well as a significant reduction in complexity.

The heart of the picoMUX system, the column module and frontend, have been validated in the 4\,K test cryostat at Princeton. 
Further development is underway to finalize the user interface and package the system in its final form.
Low-noise testing is planned with a full focal plane including TESs in a 100\,\unit{\milli\kelvin} optical cryostat.

The picoMUX system's development is closely tied to the Taurus CMB experiment as there are few viable low-power TDM systems available for stratospheric ballooning.
The combination of low digital power, and rapid firmware iteration provided by the microcontroller architecture makes picoMUX a promising warm-readout platform for Taurus and other future CMB instruments.

\appendix    

\acknowledgments 
 
Taurus is supported in the USA by NASA awards 80NSSC21K1957 and 80NSSC25K0372. 
Work at the University of Iceland is supported by the Icelandic Research Fund (Grant number: 2410656-051)

\bibliography{report} 
\bibliographystyle{spiebib} 

\end{document}